\documentclass[pdflatex,sn-mathphys-num]{sn-jnl}

\usepackage{graphicx}%
\usepackage{multirow}%
\usepackage{amsmath,amssymb,amsfonts}%
\usepackage{amsthm}%
\usepackage{mathrsfs}%
\usepackage[title]{appendix}%
\usepackage{xcolor}%
\usepackage{textcomp}%
\usepackage{manyfoot}%
\usepackage{booktabs}%
\usepackage{algorithm}%
\usepackage{algorithmicx}%
\usepackage{algpseudocode}%
\usepackage{listings}%
\usepackage{amsmath}
\usepackage{multirow} 
\usepackage{gensymb}
\usepackage{graphicx}

\usepackage[caption=false,font=footnotesize]{subfig}

\theoremstyle{thmstyleone}%
\theoremstyle{thmstyletwo}%

\theoremstyle{thmstylethree}%

\begin{document}

\title{
Evaluation of Power-Clock Waveforms for Positive Feedback Adiabatic Logic in 16\,nm FinFET Technology}


\author[1]{\fnm{Maciej Szymon} \sur{Pyrzowski}}\email{macpyrz@gmail.com}

\author[1]{\fnm{Franciszek} \sur{Łukowski}}\email{franeczek.lukowski@gmail.com}

\author*[1]{\fnm{Aida} \sur{Todri-Sanial}}\email{a.todri.sanial@tue.nl}

\affil[1]{\orgdiv{NanoComputing Research Lab, Electrical Engineering Department},
\orgname{Eindhoven University of Technology},
\orgaddress{\city{Eindhoven}, \country{The Netherlands}}}

\abstract{
Adiabatic logic can recover part of the energy stored on load capacitances through quasi-reversible switching, but its waveform-optimized operation in FinFET technology and at multi-GHz frequencies remains underexplored. This work investigates Positive Feedback Adiabatic Logic (PFAL) in the TSMC 16\,nm FinFET process. A functionally complete PFAL gate library is designed, verified, and characterised using energy-delay product optimization over power-clock amplitude, frequency, and waveform shape. The power-clock sweep shows that the minimum-energy waveform approaches a triangular shape rather than a conventional trapezoid. The optimized single-gate PFAL cells achieve up to $3.83\times$ lower energy than equivalent static CMOS gates, while sinusoidal excitation improves energy by up to $1.32\times$ compared with trapezoidal excitation and extends the valid operating range. The library is then used to construct larger combinational circuits, including a 2:1 multiplexer, a 4-bit ripple carry adder, and a 4-bit Brent--Kung carry look-ahead adder. The carry look-ahead adder reaches a gain of up to $5.3\times$ compared with the static CMOS energy estimate for the triangular power-clock.}


\maketitle

\section{Introduction} \label{sec:1}

Energy dissipation has become one of the main
limitations in modern digital integrated circuit design. Conventional
static CMOS circuits dissipate dynamic energy whenever capacitive nodes
are charged and discharged, while leakage becomes increasingly relevant
in deeply scaled technologies. Further supply-voltage scaling is increasingly constrained in deeply scaled CMOS technologies, motivating alternative low-energy circuit techniques such as adiabatic logic~\cite{teichmann2012}.

Adiabatic logic addresses this problem by replacing the fixed DC supply
with a time-varying power-clock. The power-clock acts both as
the energy source and as the timing reference of the circuit. If charge
transfer is performed slowly enough, part of the energy stored on the
load capacitance can be recovered back into the supply instead of being
fully dissipated as heat~\cite{teichmann2012, blotti2002}. Therefore,
the energy dissipation of an adiabatic transition depends not only on
capacitance and voltage, but also on the charging-path resistance and the
available transition time.

This work focuses on Positive Feedback Adiabatic Logic (PFAL), which
uses a cross-coupled latch and complementary n-channel logic networks to
generate dual-rail outputs and support partial energy recovery through
the power-clock. Adiabatic and charge-recovery logic families have previously been shown to reduce energy dissipation under suitable operating conditions~\cite{kramer1995,moon1996,teichmann2012}, while PFAL specifically has been studied for improved robustness through positive feedback and optimized transistor sizing~\cite{vetuli1996,fischer2003,blotti2002}.
However, its behaviour in advanced FinFET technology is not trivial. Lower threshold voltages can improve charge recovery, but they can also increase leakage. Higher frequencies reduce leakage time, but they also increase non-ideal adiabatic losses.

This work investigates the question of how PFAL can be designed, optimized, and scaled for energy-efficient operation in TSMC 16\,nm FinFET technology. The study focuses on gate-level efficiency, power-clock waveform selection, and arithmetic-circuit scaling. The hypothesis of this work is that waveform-optimized PFAL can achieve lower energy than equivalent static CMOS in 16\,nm FinFET technology when the power-clock waveform and circuit architecture are matched to the gate complexity. To answer this question, this work:
(i) designs and verifies a functionally complete PFAL gate library;
(ii) characterises PFAL energy behaviour over supply voltage, frequency,
and power-clock waveform shape; (iii) evaluates sinusoidal power-clock
operation as a practical alternative to trapezoidal excitation;
(iv) implements larger PFAL combinational circuits, including a
multiplexer, RCA, and CLA.


\section{Adiabatic Switching Principle and Logic} \label{sec:2}

\subsection{Adiabatic Switching}

At the switching level, the difference between conventional CMOS and adiabatic operation can be described using an equivalent load capacitance $C_{\mathrm{eq}}$ charged to a supply voltage $V_{\mathrm{DD}}$. In conventional CMOS, this capacitance is charged from a fixed supply and the stored energy is later dissipated during discharge. For a complete charge and discharge switching cycle, the dynamic switching energy is approximately

\begin{equation}
\label{eq:E_cmos}
E_{\mathrm{CMOS,cycle}} \approx C_{\mathrm{eq}} V_{\mathrm{DD}}^{2}.
\end{equation}
\\

In adiabatic switching, the same capacitance is charged through a slowly varying supply. This reduces the voltage drop across the conducting device during charge transfer, so the dissipated energy also depends on the charging-path resistance and the available transition time. If both the charging and recovery phases are considered, the adiabatic switching loss can be approximated as

\begin{equation}\label{eq:Eadia}
E_{\mathrm{adia}} = 2\,\frac{R_{\mathrm{eq}}C_{\mathrm{eq}}}{T_{\mathrm{ramp}}}\,
C_{\mathrm{eq}}V_{\mathrm{DD}}^{2} \sim \frac{1}{T_{\mathrm{ramp}}}.
\end{equation}

Here, $R_{\mathrm{eq}}$ represents the equivalent resistance of the conducting
path, $T_{\mathrm{ramp}}$ is the ramp time, and the factor of two accounts for
the charge and recovery phases. Equations~\eqref{eq:E_cmos}
and~\eqref{eq:Eadia} show that CMOS switching loss is mainly proportional to
$C_{\mathrm{eq}}V_{\mathrm{DD}}^{2}$, while adiabatic switching loss decreases
when the transition time $T_{\mathrm{ramp}}$ is increased. This expression
treats $R_{\mathrm{eq}}$ as constant, although in a transistor-level PFAL gate
the effective resistance varies during the power-clock ramp. Arbitrarily slow
operation is not optimal, since leakage energy becomes more significant at
lower frequency~\cite{jeanniot2018,teichmann2012}.

\begin{figure}
\centering
\subfloat[\label{fig:PFALgate}]{
\includegraphics[width=0.48\columnwidth]{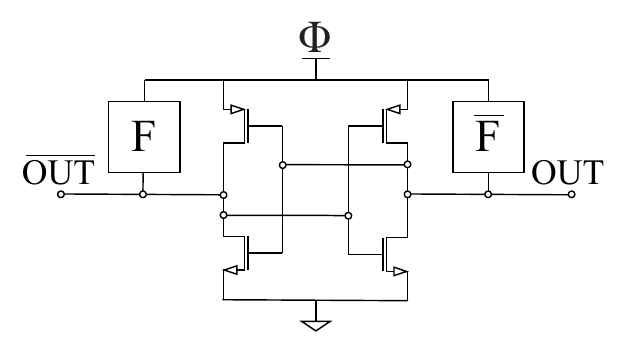}
}
\hfil
\subfloat[\label{fig:clock}]{
\includegraphics[width=0.42\columnwidth]{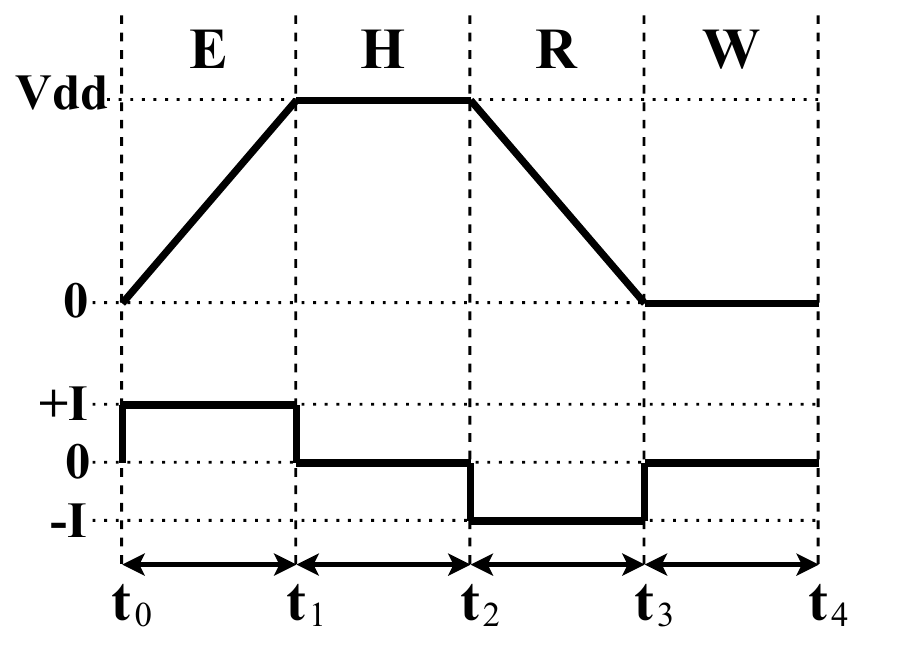}
}

\caption{PFAL logic overview: (a) general PFAL gate structure and (b) example power-clock signal.}

\label{fig}
\end{figure}

\subsection{Adiabatic Logic}
\label{sec: Adiabatic Logic}

Adiabatic logic applies the adiabatic switching principle to digital gates by replacing the fixed DC supply with a time-varying power-clock, which acts both as the energy source and the timing reference of the circuit. The general PFAL gate structure is shown in Fig.~\ref{fig:PFALgate}, and an example power-clock waveform is shown in Fig.~\ref{fig:clock}. The power-clock consists of four phases: evaluation, hold, recovery, and wait. During evaluation, the selected output node charges as the power-clock rises. During hold, the output remains valid for the next logic stage. During recovery, the power-clock falls and the current reverses direction, transferring part of the stored charge back to the power-clock supply. During wait, the gate remains inactive before the next evaluation cycle.

A logic `1' in adiabatic logic is represented by an output node that follows the power-clock waveform, while a logic `0' remains close to ground. Therefore, the output is valid only during a specific part of the power-clock period. Larger PFAL circuits require four phase-shifted power-clocks, so that each gate receives valid inputs before its evaluation phase. If the power-clock is too fast, the output nodes cannot charge and recover correctly. If it is too slow, leakage losses become more significant.

PFAL uses dual-rail signalling, meaning that each gate generates both the true and complementary outputs. This increases the number of signals compared with static CMOS, but it provides complementary outputs directly, which are required by subsequent PFAL stages.


\section{PFAL Gate Library}
\label{sec:gate_library}
\begin{figure}
\centering
\includegraphics[width=0.8\columnwidth]{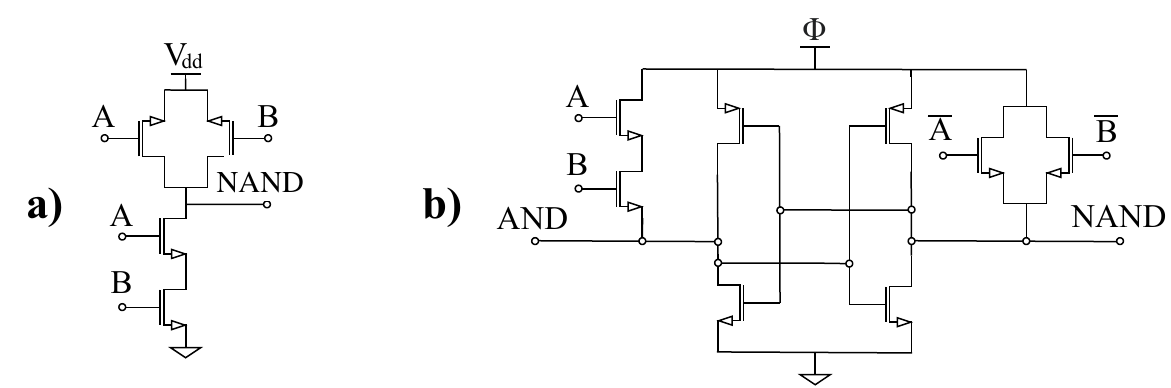}

\caption{Comparison between (a) a conventional static CMOS NAND gate and (b) the corresponding PFAL AND/NAND gate structure. 
}

\label{fig:cmos_to_pfal_nand}
\end{figure}
The implemented PFAL gate library forms the basis for the arithmetic and combinational blocks developed in this work. The library includes Buffer/NOT, AND/NAND, OR/NOR, and XOR/XNOR functionality. The presence of inversion and NAND-based logic makes the library functionally complete, while the XOR/XNOR gate provides an arithmetic building block for combinational circuits.

The construction principle is illustrated in Fig.~\ref{fig:cmos_to_pfal_nand} for the NAND gate. Each PFAL gate follows the general structure introduced in Fig.~\ref{fig:PFALgate}: a cross-coupled latch is combined with two complementary NMOS logic trees that implement $F$ and $\overline{F}$. One network follows the pull-down structure of the equivalent static-CMOS function, while the complementary network follows the corresponding pull-up topology with PMOS devices replaced by NMOS devices driven by complementary inputs. These networks, together with the cross-coupled latch, generate the true and complementary PFAL outputs.

The AND/NAND gate, shown in Fig.~\ref{fig:cmos_to_pfal_nand}, is a versatile PFAL cell because the same topology can also implement OR/NOR using De Morgan's law. If the complementary inputs $\overline{A}$ and $\overline{B}$ are applied to the physical AND/NAND gate, the NAND output behaves as OR, while the AND output behaves as NOR. Therefore, a separate OR/NOR schematic is not required.

\begin{figure}
\centering
\subfloat[PFAL Buffer/NOT gate.\label{fig:inv_pfal}]{
\includegraphics[width=0.38\columnwidth]{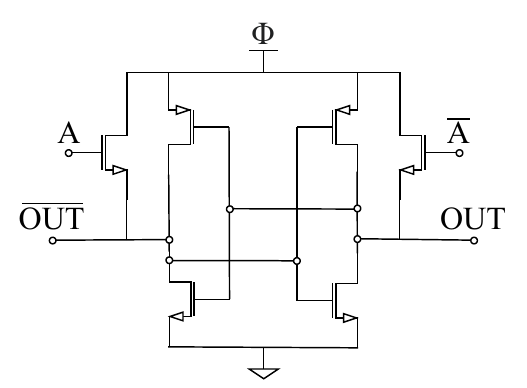}
}
\hfil
\subfloat[PFAL AND/NAND gate.\label{fig:and_nand_pfal}]{
\includegraphics[width=0.5\columnwidth]{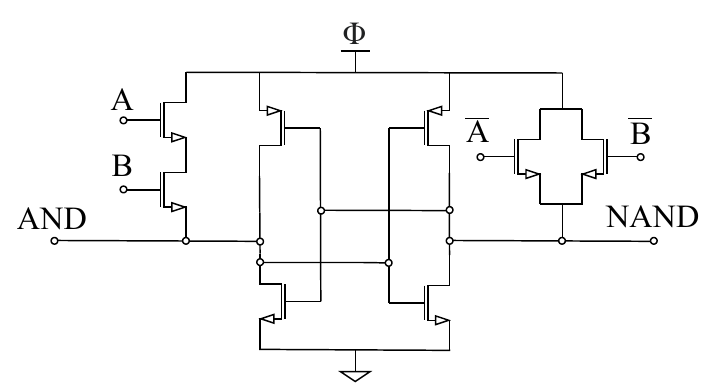}
}
\hfil
\subfloat[PFAL XOR/XNOR gate.\label{fig:xor_xnor_pfal}]{
\includegraphics[width=0.55\columnwidth]{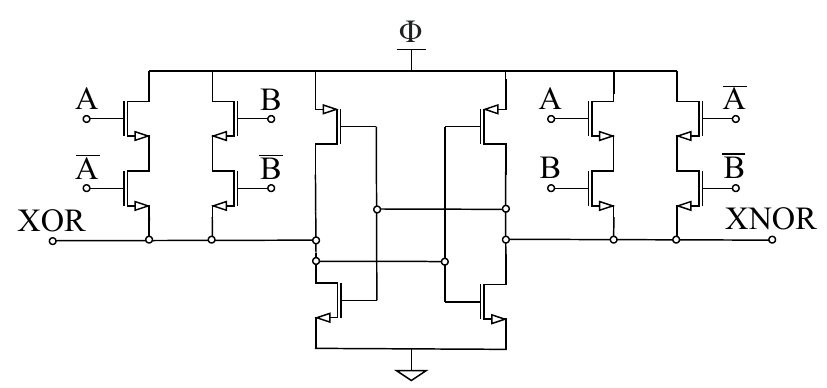}
}

\caption{Implemented PFAL gate structures: Buffer/NOT, AND/NAND, and XOR/XNOR cells.}

\label{fig:basic_pfal_gates}
\end{figure}

The implemented gate structures are shown in Fig.~\ref{fig:basic_pfal_gates}. The Buffer/NOT gate in Fig.~\ref{fig:inv_pfal} is the simplest PFAL cell and is mainly used to generate complementary signals or to delay signals for phase alignment. The XOR/XNOR gate in Fig.~\ref{fig:xor_xnor_pfal} requires a larger complementary NMOS logic tree, making it the most complex gate in the basic PFAL library. It is therefore used to evaluate how gate complexity affects PFAL operation and also serves as an arithmetic building block for adders and multiplexer-based circuits.

\begin{table}
\centering
\caption{Transistor-count comparison between the implemented PFAL gate library and representative equivalent static CMOS implementations. The CMOS counts assume that both true and complementary outputs are generated for fair comparison with the dual-rail PFAL gates.}

\label{tab:gate_summary}
\begin{tabular}{|l|ccc|ccc|}
\hline
 & \multicolumn{3}{c|}{\textbf{PFAL}}
 & \multicolumn{3}{c|}{\textbf{CMOS}} \\
\hline
\textbf{Gate}
 & \textbf{PMOS} & \textbf{NMOS} & \textbf{Total}
 & \textbf{PMOS} & \textbf{NMOS} & \textbf{Total} \\
\hline
Buffer/NOT & 2 & 4  & 6  & 2 & 2 & 4  \\
AND/NAND   & 2 & 6  & 8  & 3 & 3 & 6  \\
XOR/XNOR   & 2 & 10 & 12 & 6 & 6 & 12 \\
\hline
\end{tabular}
\end{table}

The transistor-count comparison is given in Table~\ref{tab:gate_summary}. The listed PFAL transistor counts include the cross-coupled latch and the complementary NMOS functional networks required to generate the true and complementary outputs.

\begin{figure}[!t]
\centering
\includegraphics[width=0.6\linewidth]{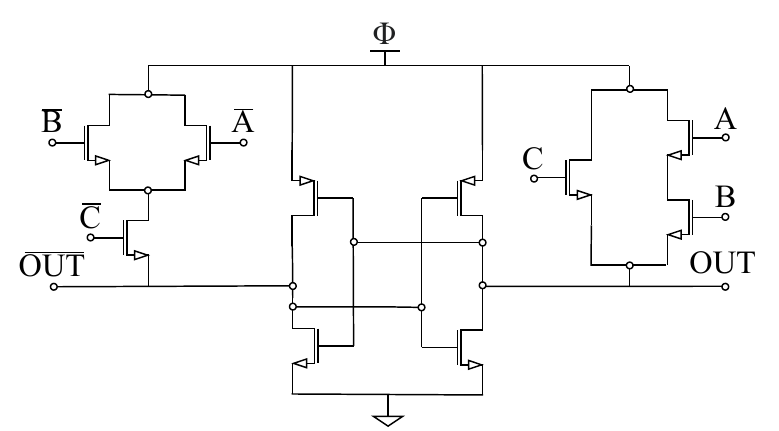}

\caption{Implemented PFAL AO21 gate used in the Brent--Kung carry look-ahead adder.}

\label{fig:ao21_pfal}
\end{figure}

The AO21 gate used later in the Brent--Kung carry look-ahead adder is shown in Fig.~\ref{fig:ao21_pfal}. It implements the group-generate term of the prefix operation and allows the CLA to be built from smaller PFAL cells instead of larger full-adder blocks.


\section{Power-Clock Shape Optimization}
\label{sec:clock_optimization}
After the basic PFAL gate library had been implemented, the influence of the power-clock waveform on gate-level energy dissipation was investigated. Since the PFAL output is directly driven by the power-clock, the waveform shape affects both the charging loss during evaluation and the recovered energy during the falling transition. In the adiabatic logic studies reviewed in this work, the implemented power-clock waveform is consistently assumed to be trapezoidal, as shown in Fig.~\ref{fig:clock}. Therefore, the trapezoidal waveform was used as the starting point for the power-clock optimization.

\subsection{Sweep Definition}
\label{subsec:clock_sweep}
To study the effect of the waveform shape, the trapezoidal power-clock was swept by varying the relative durations of the evaluation, hold, recovery, and wait phases while keeping the total clock period $T$ fixed. The clock pulse width was assigned to both the hold and wait phases, so $t_{\mathrm{wait}}=t_{\mathrm{hold}}$. The clock rise time represented the evaluation phase, and the remaining time in the period was assigned to the recovery phase:

\begin{equation}
t_{\mathrm{rec}} = T - t_{\mathrm{wait}} - t_{\mathrm{hold}} - t_{\mathrm{eval}} .
\label{eq:trec}
\end{equation}
Here, $t_{\mathrm{eval}}$ is the evaluation time, $t_{\mathrm{hold}}$ is the hold time, $t_{\mathrm{rec}}$ is the recovery time, and $t_{\mathrm{wait}}$ is the wait time. The sweep was performed for the Buffer/NOT, AND/NAND, and XOR/XNOR PFAL gates to determine whether the minimum-energy waveform was gate-dependent or whether a common waveform shape could be used for the full library.

The energy drawn from the power-clock was obtained by integrating the instantaneous power over the relevant simulation window,

\begin{equation}
E = \int_{t_0}^{t_1} V_{\mathrm{CLK}}(t) I_{\mathrm{CLK}}(t)\ \mathrm{d}t .
\label{eq:energy}
\end{equation}
The energy values were extracted in Cadence according to Eq.~\eqref{eq:energy}. For the Buffer/NOT gate, the integration covered two clock pulses corresponding to input logic `0' and logic `1'. For the AND/NAND and XOR/XNOR gates, the integration covered the four possible input combinations.

\subsection{Sweep Results}
\label{subsec:clock_results}
\begin{figure}
\centering
\includegraphics[width=0.8\columnwidth]{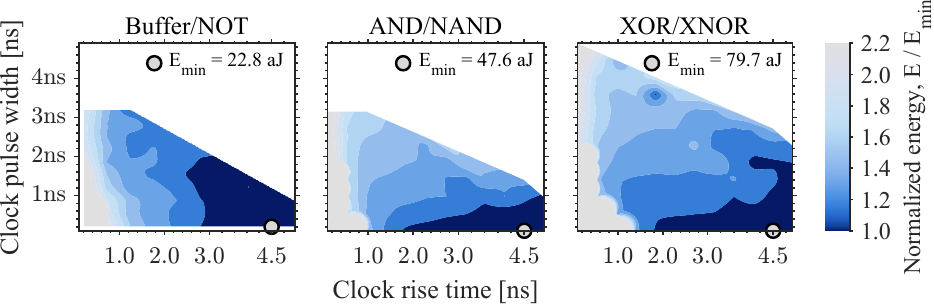}

\caption{Clock parameter sweep of the Buffer/NOT, AND/NAND, and XOR/XNOR PFAL gates.}

\label{fig:clock_sweep}
\end{figure}
\begin{figure}
\centering
\includegraphics[width=0.75\columnwidth]{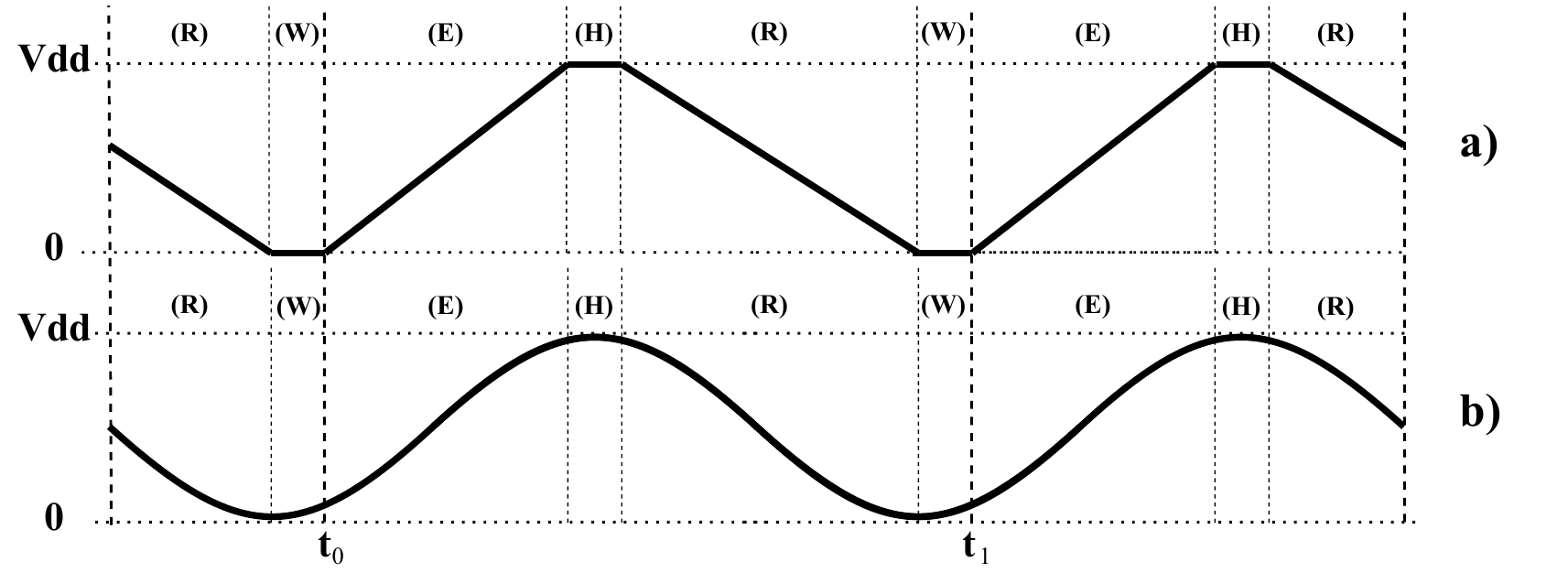}

\caption{Power-clock waveforms: (a) triangular signal, and (b) sinusoidal signal.}

\label{fig:clocksignal}
\end{figure}

The results of the parameter sweep are shown in Fig.~\ref{fig:clock_sweep}. The heatmap visualization shows that, for all investigated PFAL gates, the lowest energy consumption occurs when the hold and wait intervals are short and the transition time is long. Therefore, the optimum trapezoidal waveform does not correspond to a conventional trapezoid with a long flat region, but instead approaches the triangular power-clock shown in Fig.~\ref{fig:clocksignal}a. This agrees with the adiabatic switching principle, since a slower voltage transition reduces the voltage drop across the conducting devices during charge transfer. The sweep also shows that the optimized gates operate at attojoule-level energy values.

\begin{table}[!t]
\centering
\caption{Minimum-energy points obtained from the trapezoidal power-clock sweep. The optimum waveform approaches a triangular power-clock for all investigated gates.}

\label{tab:clock_sweep_minima}
\begin{tabular}{| l | c | c |}
\hline
\textbf{PFAL gate} & \textbf{Minimum energy} & \textbf{Reduction vs. CMOS} \\
\hline
Buffer/NOT & $22.77~\mathrm{aJ}$ & $3.65\times$ \\
AND/NAND   & $45.93~\mathrm{aJ}$ & $3.83\times$ \\
XOR/XNOR   & $78.41~\mathrm{aJ}$ & $3.71\times$ \\
\hline
\end{tabular}

\end{table}

The corresponding minimum-energy points are summarized in Table~\ref{tab:clock_sweep_minima}. These values correspond to the final optimized operating points obtained after the initial clock-parameter sweep. The simulations used buffered PFAL inputs generated by preceding Buffer/NOT stages, so the applied input waveform was a realistic adiabatic waveform rather than an ideal voltage source. This buffered-input setup gave slightly lower energy values than the initial sweep-level results. Therefore, the table reports the final optimized energies, while Fig.~\ref{fig:clock_sweep} shows the sweep-level trend used to identify the near-triangular optimum region.

Based on this observation, a sinusoidal power-clock was also investigated. The sinusoidal waveform, shown in Fig.~\ref{fig:clocksignal}b, is similar to the triangular waveform because it avoids abrupt voltage steps and does not contain a long constant-voltage hold phase. More importantly, it is more practical for physical implementation, since it can be generated by an LC resonant oscillator. The sinusoidal clock therefore represents a compromise between the minimum-energy waveform found in the sweep and a realistically implementable power-clock source.

The sinusoidal power-clock was evaluated at a supply amplitude of $400\,\mathrm{mV}$ and an operating frequency of $100\,\mathrm{MHz}$. Under these conditions, the Buffer/NOT gate showed almost the same energy efficiency as in the optimized trapezoidal case, the AND/NAND gate consumed approximately $20\%$ more energy, and the XOR/XNOR gate consumed approximately $9.4\%$ more energy. The sinusoidal waveform remains practical due to its compatibility with LC resonant oscillator generation, even when it is not the absolute minimum-energy solution.

\section{Combinational Circuits}
\label{sec:comb_circuits}

The gate library and waveform optimization of the preceding sections establish the foundation for constructing larger PFAL circuits. In this work, three combinational building blocks were implemented and verified in TSMC 16\,nm: a 2:1 multiplexer, a 4-bit ripple carry adder (RCA), and a 4-bit carry look-ahead adder (CLA). These blocks show that the PFAL standard-cell approach extends from single gates to multi-stage arithmetic logic.

\subsection{Circuit-Level Building Blocks}
\label{subsec:comb_building_blocks}

\begin{table}[!t]
\centering
\caption{Input-node mapping used to implement the PFAL 2:1 multiplexer by reusing the XOR/XNOR topology.}

\label{tab:xor_mux_mapping}
\begin{tabular}{|c|c|}
\hline
\textbf{XOR node} & \textbf{MUX node} \\
\hline
$A$ & $A$ \\
$\overline{A}$ & $B$ \\
$B$ & $\overline{S}$ \\
$\overline{B}$ & $S$ \\
\hline
\end{tabular}

\end{table}

\begin{figure}[!t]
\centering
\includegraphics[width=0.9\linewidth]{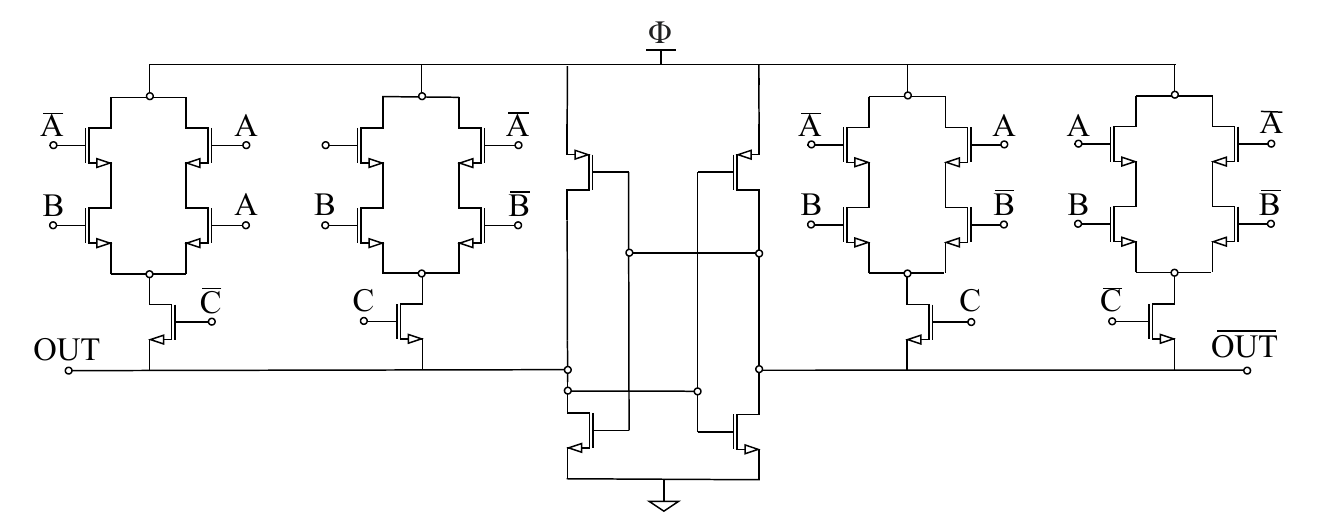}

\caption{PFAL 1-bit sum gate used in the ripple carry adder.}

\label{fig:sum_gate}
\end{figure}

\begin{figure}[!t]
\centering
\includegraphics[width=0.75\columnwidth]{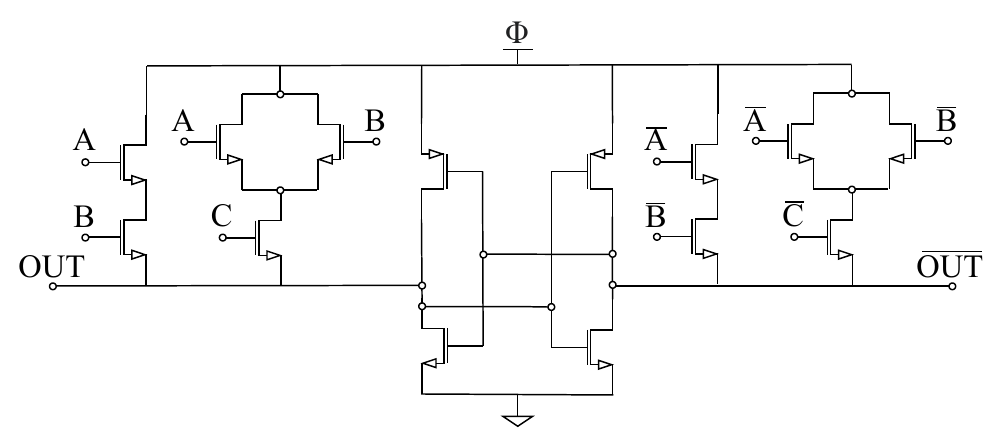}

\caption{PFAL 1-bit carry gate used in the ripple carry adder.}

\label{fig:carry_gate}
\end{figure}

\begin{figure}[!t]
\centering
\subfloat[4-bit PFAL ripple\\ carry adder.\label{fig:rca_arch}]{
\includegraphics[width=0.37\linewidth]{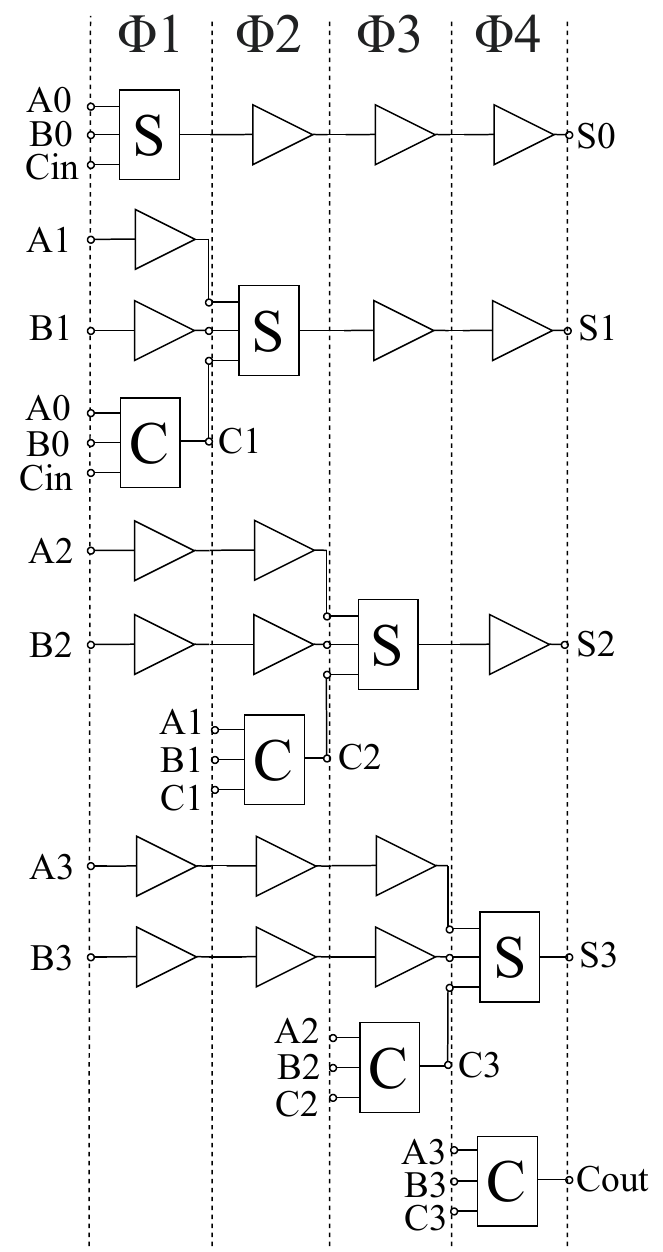}
}
\hfill
\subfloat[4-bit PFAL Brent--Kung carry look-ahead adder.\label{fig:cla_arch}]{
\includegraphics[width=0.5\linewidth]{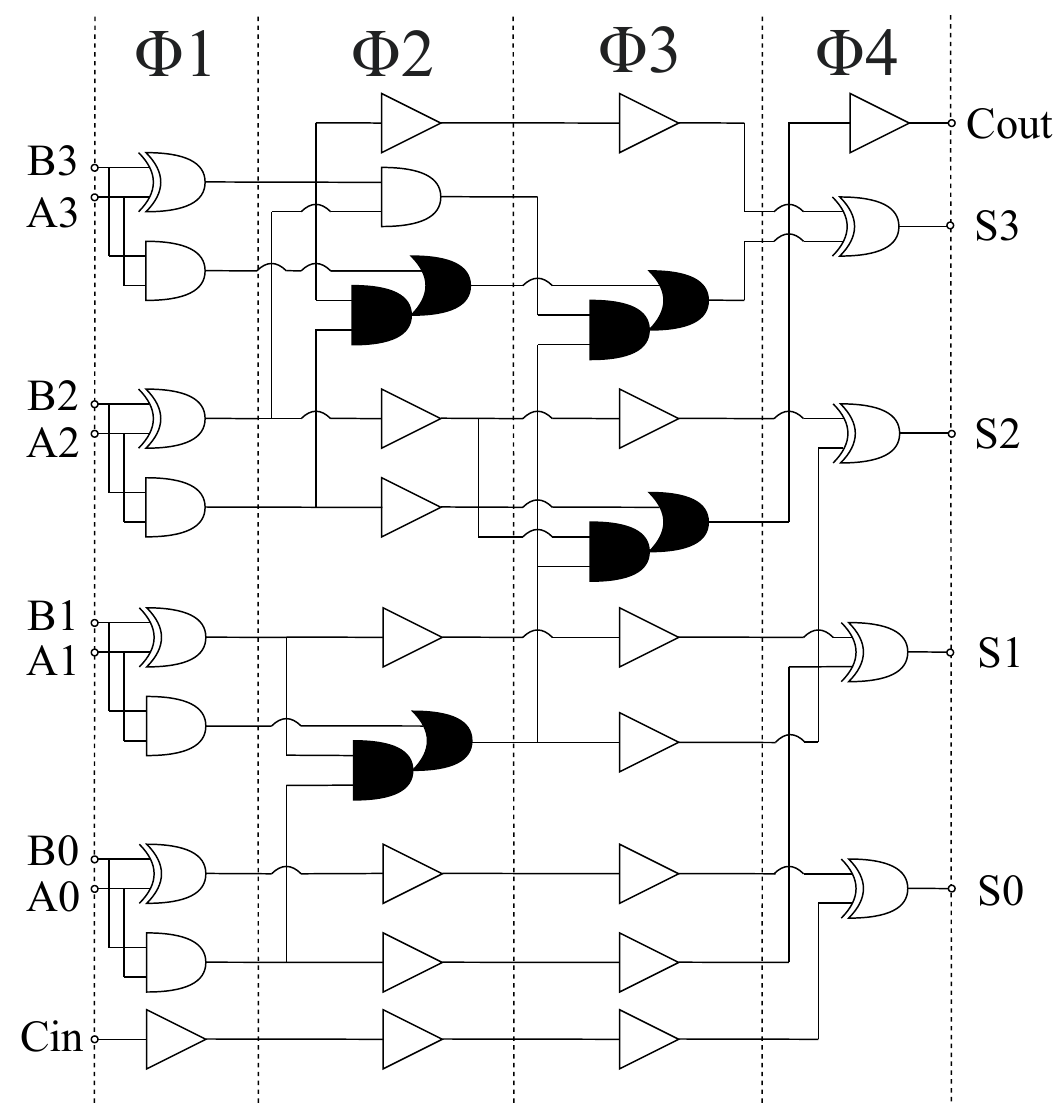}
}

\caption{Implemented PFAL adder architectures: (a) 4-bit ripple carry adder, where the carry propagates sequentially from the least significant bit to the most significant bit, and (b) 4-bit Brent--Kung carry look-ahead adder, where the prefix structure computes carries using smaller logic cells and phase-alignment buffers.}

\label{fig:adder_architectures}
\end{figure}

The circuit-level building blocks used to construct the three combinational circuits are collected here. The input-node mapping used to implement the PFAL 2:1 multiplexer by reusing the XOR/XNOR topology is given in Table~\ref{tab:xor_mux_mapping}. The PFAL sum and carry gates used in the ripple carry adder are shown in Fig.~\ref{fig:sum_gate} and Fig.~\ref{fig:carry_gate}, respectively, and the two implemented adder architectures are shown in Fig.~\ref{fig:adder_architectures}.

\subsection{2:1 Multiplexer}

A 2:1 multiplexer was implemented as one of the basic combinational PFAL cells. The multiplexer selects one of two input signals, $A$ or $B$, depending on the value of the select signal. When the select signal is low, the multiplexer forwards input $A$ to the output. When the select signal is high, the multiplexer forwards input $B$ to the output.

The implemented PFAL multiplexer uses the same transistor-level topology as the previously designed XOR/XNOR gate. Only the input node assignments are changed, and the selected output polarity is inverted. Therefore, no additional dedicated PFAL topology is required for the multiplexer. The mapping between the XOR/XNOR gate inputs and the multiplexer inputs is shown in Table~\ref{tab:xor_mux_mapping}.

\subsection{4-Bit Ripple Carry Adder}
The 4-bit RCA shown in Fig.~\ref{fig:rca_arch} is built from four cascaded PFAL 1-bit full-adder stages. The sum and carry gate architectures were based on the PFAL full-adder structures reported in~\cite{dillikumar2013} and adapted to the developed PFAL library. Each stage receives one bit pair $(A_i,B_i)$ and the carry from the previous bit position, and consists of a PFAL sum gate and carry gate, shown in Fig.~\ref{fig:sum_gate} and Fig.~\ref{fig:carry_gate}.

The carry output of bit $i$ must become valid before bit $i+1$ can evaluate its own carry, so the carry path is inherently sequential and determines the critical path. Since PFAL evaluates one logic level per power-clock phase, each carry stage requires one adiabatic phase, giving approximately one phase of latency per adder bit. The sum bit of each stage is generated in the same phase as the corresponding carry calculation. Because different bit positions become valid in different phases, buffer stages are inserted to align the remaining signals before they are used by later logic. This phase-alignment overhead increases the total gate count and switched capacitance of the RCA. Although the RCA has a simple and regular structure, its latency therefore scales linearly with bit width.

The same structure can implement subtraction by two's-complement arithmetic. A multiplexer selects $\overline{B}$ instead of $B$, and the input carry is set to $C_{\mathrm{in}}=1$, giving $A-B=A+\overline{B}+1$. For the 4-bit implementation, $C_{\mathrm{out}}=1$ indicates a non-negative result, while $C_{\mathrm{out}}=0$ indicates a negative result whose magnitude is obtained by inverting the output bits and adding one.

\subsection{4-Bit Carry Look-Ahead Adder}
To reduce the carry-chain latency of the ripple carry adder, a 4-bit carry look-ahead adder was implemented. The use of a carry look-ahead structure for adiabatic standard-cell logic follows the approach reported by Blotti \textit{et al.}~\cite{blotti2002}, while the prefix network itself is based on the Brent--Kung adder~\cite{brent1982}. The architecture was adapted to the four-phase PFAL timing scheme and to the available PFAL standard-cell library. Instead of propagating the carry sequentially through all bit positions, the CLA first generates local propagate and generate signals and then combines them through a prefix network.

The implemented architecture is shown in Fig.~\ref{fig:cla_arch}. In the first stage, each bit generates local propagate and generate signals,

\begin{equation}
p_i = A_i \oplus B_i, \qquad g_i = A_iB_i ,
\label{eq:pg}
\end{equation}
where the propagate term is generated using a PFAL XOR gate and the generate term using a PFAL AND gate. These signals are then passed to the prefix-carry network. For two adjacent carry groups, the prefix operator combines the group-generate and group-propagate terms as

\begin{equation}
(G_i,P_i)\circ(G_j,P_j)=\left(G_i+P_iG_j,\;P_iP_j\right).
\label{eq:prefix}
\end{equation}
In the PFAL implementation, the group-generate term $G_i+P_iG_j$ is realized using an AO21 cell, shown in Fig.~\ref{fig:ao21_pfal}, while the group-propagate term $P_iP_j$ is realized using an AND2 cell. The black gates in Fig.~\ref{fig:cla_arch} are the AO21 cells used for the Brent--Kung prefix operation. Additional Buffer/NOT cells are inserted where phase alignment is required. After the carry signals are generated by the prefix tree, the final sum bits are computed using XOR gates according to

\begin{equation}
S_i = p_i \oplus C_i ,
\label{eq:sum}
\end{equation}
where $C_i$ is the carry into bit position $i$. Therefore, the complete 4-bit CLA consists of an initial propagate/generate stage, a Brent--Kung prefix-carry network, phase-alignment buffers, and a final sum-generation stage. The implemented circuit uses only PFAL Buffer/NOT, AND2, XOR2, and AO21 cells, keeping the architecture compatible with the developed PFAL gate library.

The implemented 4-bit CLA evaluates addition using the four available PFAL power-clock phases, because it was designed for $C_{\mathrm{in}}=0$. Therefore, this exact implementation does not directly support subtraction. To use the same architecture for subtraction, $C_{\mathrm{in}}$ would need to be set to one and included in the prefix-carry computation, which would introduce an additional power-clock phase and extra phase-alignment buffers so that the input carry can correctly influence the later carry and sum stages. With this added carry-input phase, the phase-scaling model becomes $N_{\mathrm{phases}}=\log_2(N)+3$.

\section{Results and Discussion}\label{sec:results}

This section summarizes the single-gate and combinational-circuit results. The PFAL gates are evaluated using EDP over supply voltage and frequency, the sinusoidal and trapezoidal power-clocks are compared using energy ratio, and the 4-bit RCA and CLA are compared against static CMOS using energy gain.

All transistor-level simulations were performed in Cadence Virtuoso using the TSMC 16\,nm FinFET process design kit, with transistor models based on BSIM-CMG. For the single-gate characterisation, each PFAL device under test was driven by preceding PFAL Buffer/NOT stages clocked by earlier power-clock phases, so that the input waveform resembled a realistic adiabatic signal rather than an ideal voltage source. The output was loaded with a minimum-sized CMOS inverter and an output capacitor.

The energy drawn from the power-clock was obtained according to Eq.~\eqref{eq:energy}. For gate-level simulations, the integration window covered a complete set of input combinations and the result was normalised accordingly. For periodic waveform comparisons, the energy was normalised per clock period. The EDP was calculated as $\mathrm{EDP}=E/f_{\mathrm{CLK}}$, while energy gain was used only for comparisons against static CMOS.

The outputs were verified against their expected Boolean values. An operating point was considered invalid if the logic state was incorrect or if residual distortion could corrupt the next evaluation. In the EDP and energy-gain plots, missing or clipped regions therefore indicate failed operating points and define the valid operating region.

For the RCA and CLA comparison, the CMOS reference was architecture-matched. The PFAL RCA was compared with an equivalent static-CMOS RCA estimate, while the PFAL CLA was compared with an equivalent static-CMOS Brent--Kung CLA estimate. In both cases, the CMOS reference energy was estimated by summing the individually measured energies of the corresponding static CMOS gates. This approximation neglects inter-gate loading effects and gate-specific load variations, so it should be interpreted as an estimated CMOS reference rather than a full post-layout CMOS implementation. All PFAL and CMOS energy comparisons assume an activity factor of one.

\subsection{Single-Gate PFAL Results}

The implemented PFAL gate library is functionally complete because it includes inversion and NAND-based logic, while the dual-rail structure naturally provides both true and complementary outputs. This enables the construction of larger PFAL circuits, but at the cost of higher transistor count, complementary input requirements, and strict power-clock phase alignment.

The sinusoidal and trapezoidal power-clocks were compared using the energy ratio

\begin{equation}
G_{\mathrm{sin/trap}} = \frac{E_{\mathrm{trap}}}{E_{\mathrm{sin}}} .
\label{eq:gain_sin_trap}
\end{equation}
This comparison does not imply that sinusoidal excitation is more efficient than an ideal linear ramp. Jeanniot's $\pi^2/8$ result applies to an ideal RC charging model with constant resistance and an ideal linear reference~\cite{jeanniot2018}. In this work, the sinusoidal waveform is compared against the optimized trapezoidal PFAL implementation characterised in Section~\ref{sec:clock_optimization}. The observed gain therefore shows that sinusoidal excitation outperforms the evaluated trapezoidal PFAL implementation, not that it violates the ideal ramp limit.

\begin{figure}[htbp]
\centering
\subfloat[Buffer/NOT gate, maximum gain of $1.19$ at $V_{\mathrm{CLK}}=1.0\,\mathrm{V}$ and $f=753\,\mathrm{MHz}$.\label{fig:gain_sin_trap_inv}]{
\includegraphics[width=0.65\columnwidth]{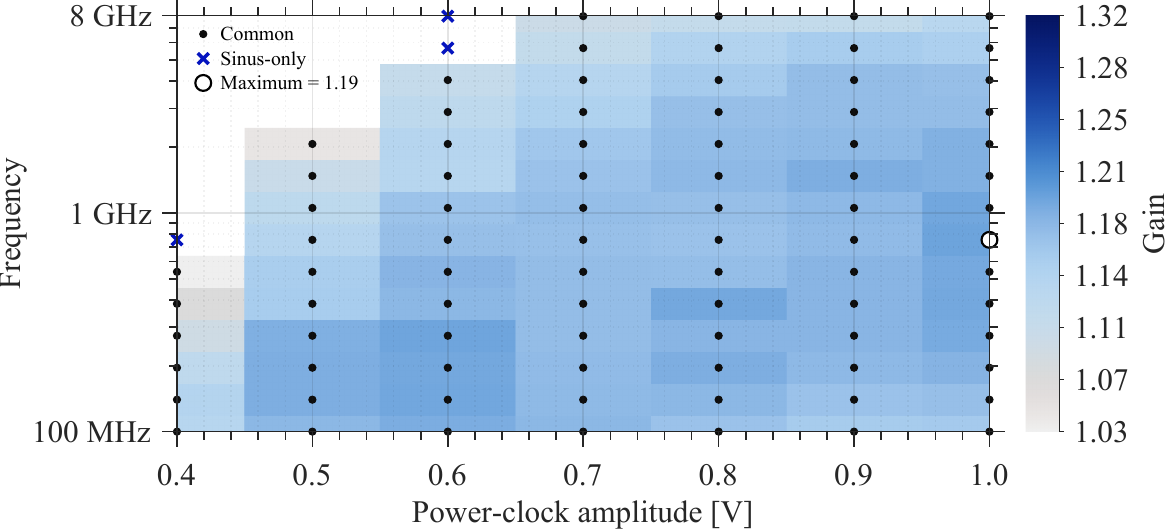}
}

\subfloat[AND/NAND gate, maximum gain of $1.18$ at $V_{\mathrm{CLK}}=0.6\,\mathrm{V}$ and $f=140\,\mathrm{MHz}$.\label{fig:gain_sin_trap_nand}]{
\includegraphics[width=0.65\columnwidth]{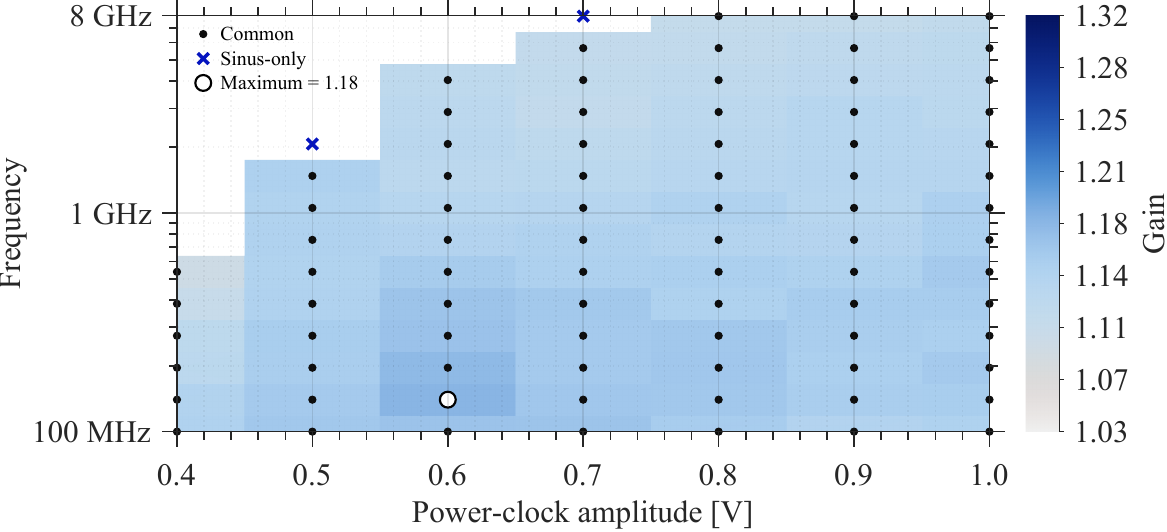}
}

\subfloat[XOR/XNOR gate, maximum gain of $1.32$ at $V_{\mathrm{CLK}}=0.5\,\mathrm{V}$ and $f=384\,\mathrm{MHz}$.\label{fig:gain_sin_trap_xor}]{
\includegraphics[width=0.65\columnwidth]{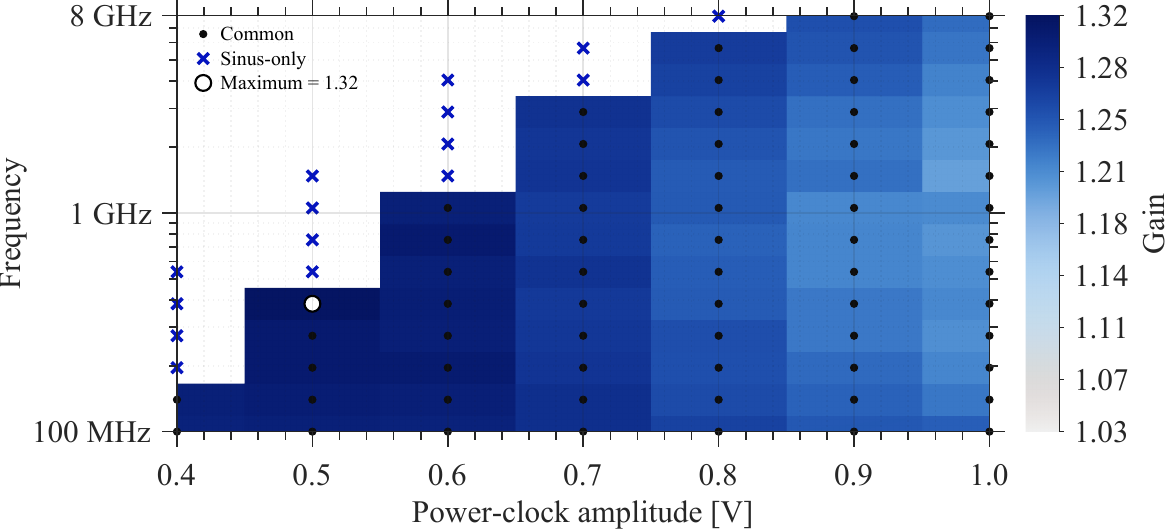}
}

\caption{Energy ratio $E_{\mathrm{trap}}/E_{\mathrm{sin}}$ for the investigated PFAL gates. A value larger than one means the sinusoidal power-clock consumes less energy at the same operating point. Cross markers indicate sinusoidal-only valid operating points.}

\label{fig:gain_sin_trap_all}
\end{figure}

\begin{figure}[htbp]
\centering
\subfloat[Buffer/NOT gate, minimum EDP of $1.28\times10^{-26}\,\mathrm{J\,s}$ at $V_{\mathrm{CLK}}=0.7\,\mathrm{V}$ and $f=7.9\,\mathrm{GHz}$.\label{fig:edp_inv}]{
\includegraphics[width=0.65\columnwidth]{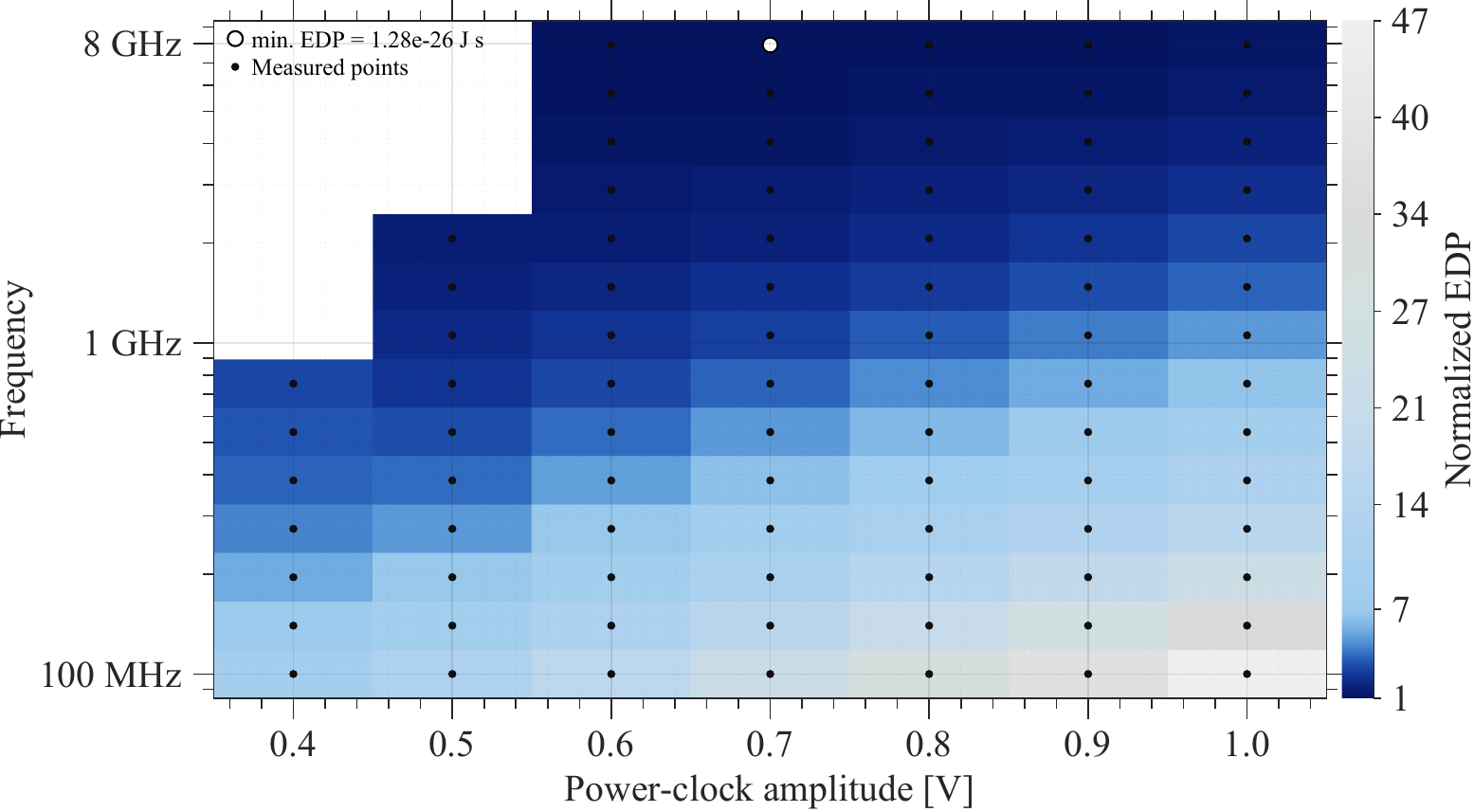}
}

\subfloat[AND/NAND/OR/NOR gate, minimum EDP of $2.60\times10^{-26}\,\mathrm{J\,s}$ at $V_{\mathrm{CLK}}=0.7\,\mathrm{V}$ and $f=7.9\,\mathrm{GHz}$.\label{fig:edp_nand}]{
\includegraphics[width=0.65\columnwidth]{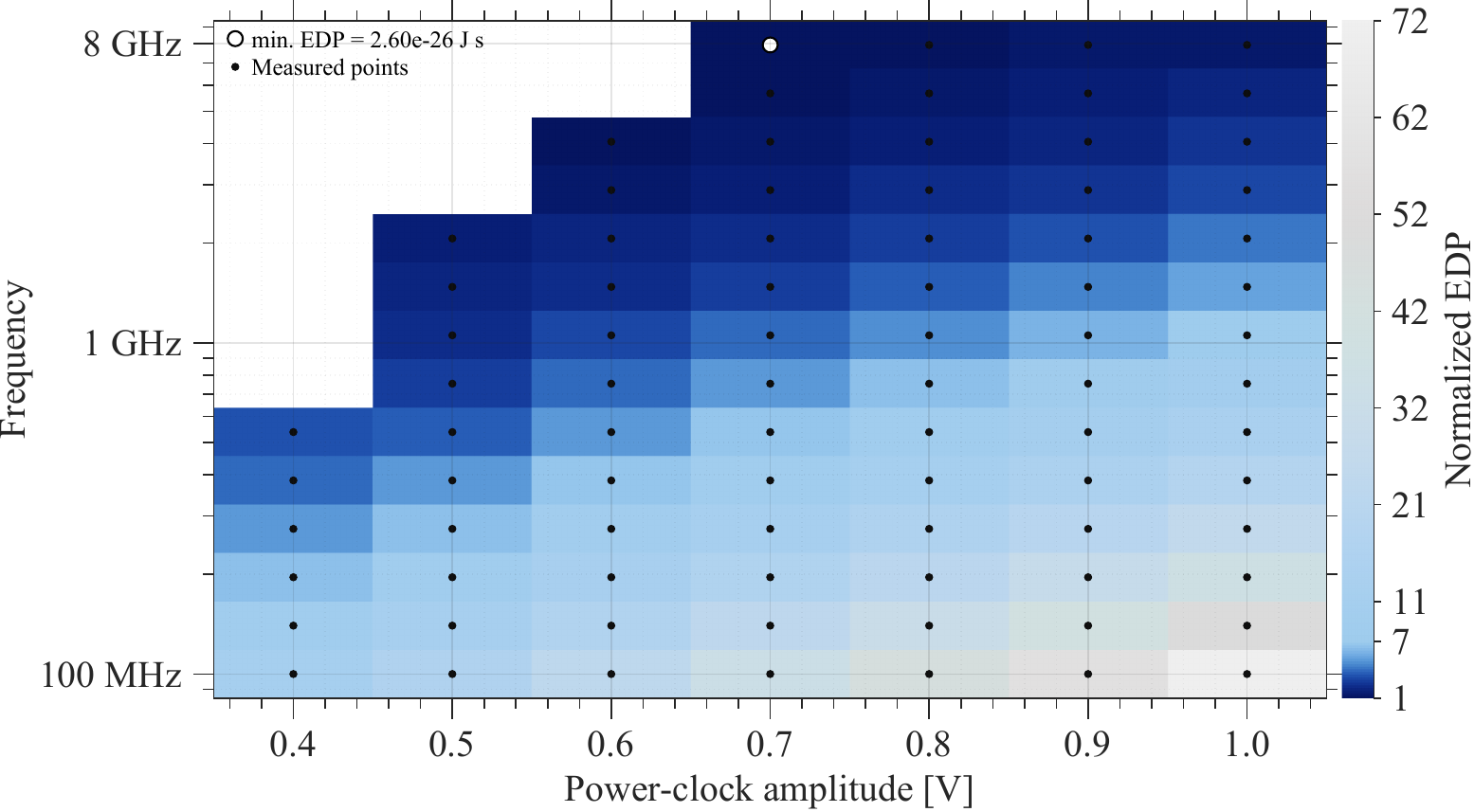}
}

\subfloat[XOR/XNOR gate, minimum EDP of $3.96\times10^{-26}\,\mathrm{J\,s}$ at $V_{\mathrm{CLK}}=0.8\,\mathrm{V}$ and $f=7.9\,\mathrm{GHz}$.\label{fig:edp_xor}]{
\includegraphics[width=0.65\columnwidth]{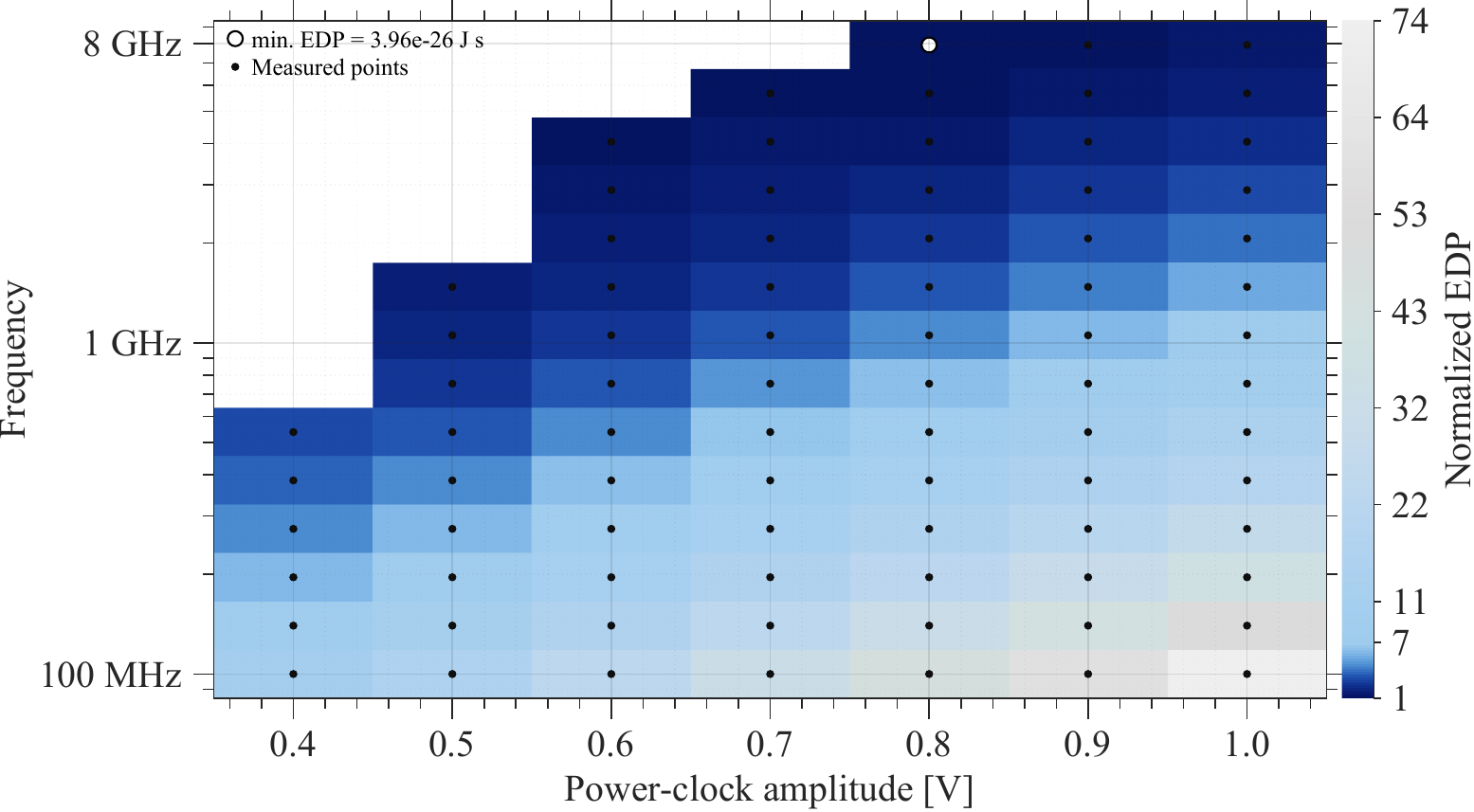}
}

\caption{Energy-delay product of the implemented PFAL gates over power-clock amplitude and operating frequency. Missing regions indicate failed operating points and define the valid operating region of each gate.}

\label{fig:edp_all}
\end{figure}

The trapezoidal PFAL implementation already achieves up to $3.83\times$ lower energy than equivalent static CMOS at its optimized operating points, as shown in Table~\ref{tab:clock_sweep_minima}. The comparison maps for the full gate library are shown in Fig.~\ref{fig:gain_sin_trap_all}, where the strongest improvement of $1.32\times$ is reached by the XOR/XNOR gate at $V_{\mathrm{CLK}}=0.5\,\mathrm{V}$ and $f=384\,\mathrm{MHz}$. The cross markers indicate sinusoidal-only operating points, where the sinusoidal waveform remains functional at lower supply voltage or higher frequency while the trapezoidal waveform fails. For all three gates, the ratio $E_{\mathrm{trap}}/E_{\mathrm{sin}}$ remained above one across the evaluated valid operating points, meaning that the sinusoidal waveform was consistently more energy efficient than the conventional trapezoidal reference. This behaviour of the XOR/XNOR gate is consistent with its more complex transistor network and stronger sensitivity to smoother charging. Overall, the sinusoidal waveform extends the usable voltage-frequency range of every gate in the library, supporting the conclusion that smoother power-clock waveforms are better suited to this 16\,nm PFAL implementation. Since sinusoidal clocks can be generated using resonant LC oscillators, this makes them a practical candidate for PFAL power-clock generation.

The single-gate energy-delay product was evaluated over power-clock amplitude and operating frequency for the Buffer/NOT, AND/NAND, and XOR/XNOR PFAL gates, as shown in Fig.~\ref{fig:edp_all}. The Buffer/NOT and AND/NAND gates reach similar minimum EDP values, while the XOR/XNOR gate reaches a higher minimum EDP close to the boundary of the correct-operation envelope. This follows from its larger transistor count and higher internal capacitance, which increase the effective charging path and make the gate harder to drive adiabatically.

All investigated gates show an optimum over power-clock amplitude and frequency, occurring at relatively high frequency and moderate power-clock amplitude. At low frequency, leakage accumulates over the longer clock period and becomes more significant as the circuit spends more time in each cycle. At excessive frequency or unsuitable power-clock amplitude, the latch nodes cannot charge and recover correctly, producing failed operating points. The minimum EDP is therefore obtained only when the power-clock is slow enough for adiabatic charging but not so slow that leakage dominates.

\subsection{Combinational Circuit Results}

\begin{figure}[htbp]
\centering
\includegraphics[width=0.7\columnwidth]{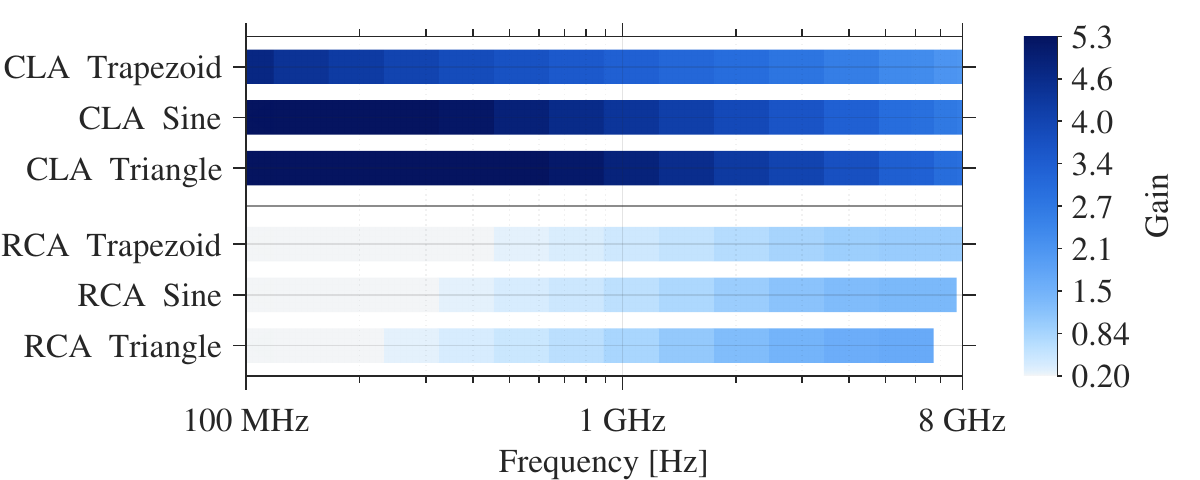}

\caption{Energy gain $E_{\mathrm{CMOS}}/E_{\mathrm{PFAL}}$ of the 4-bit PFAL CLA and RCA at $V_{\mathrm{CLK}}=900\,\mathrm{mV}$ for triangular, sinusoidal, and trapezoidal power-clocks. Missing or clipped regions indicate failed PFAL operating points.}

\label{fig:clarca_gain}
\end{figure}

The larger PFAL circuits were evaluated to determine how the single-gate behaviour scales to multi-stage arithmetic logic. The combinational circuit results are shown in Fig.~\ref{fig:clarca_gain}. For the RCA and CLA comparison, the energy gain relative to static CMOS is defined as $G_E = E_{\mathrm{CMOS}}/E_{\mathrm{PFAL}}$.

The 4-bit CLA and 4-bit RCA were compared against equivalent static CMOS adders using this metric. Both adders can achieve energy gain over CMOS. In particular, the CLA reaches a gain of up to $5.3\times$ compared with the static CMOS energy estimate for the triangular power-clock. The CLA also performs better overall because it achieves higher gain across the investigated waveforms and remains functional over a wider frequency range. The RCA degrades more strongly as frequency increases. Blank regions in the gain map mark the functional limit of each PFAL adder.

This difference is caused by the gate-level structure of the two adder architectures. The RCA is built from full-adder sum and carry blocks, which contain relatively large transistor networks. These larger PFAL cells have higher internal capacitance and longer effective charging paths, making it harder for the power-clock to charge and recover the latch nodes correctly at high frequency. In contrast, the Brent--Kung CLA is built from smaller cells such as Buffer/NOT, AND, XOR, and AO21 gates. Although the CLA contains more gates at the architectural level, each individual PFAL gate is easier to drive with the power-clock.

\begin{figure}[htbp]
\centering
\includegraphics[width=0.7\linewidth]{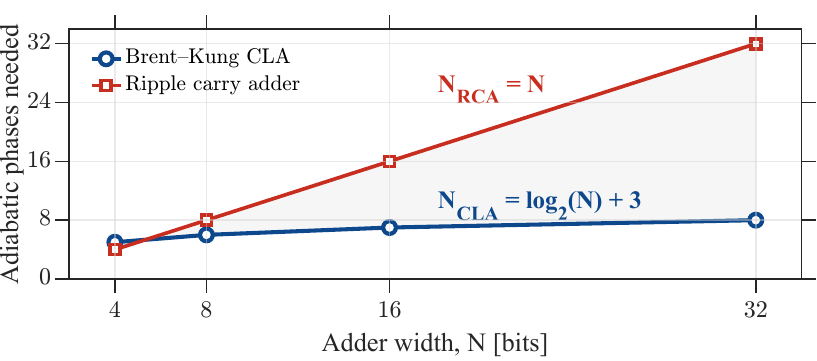}

\caption{Adiabatic phase scaling of PFAL adders. The ripple carry adder scales approximately linearly with bit width, while the Brent--Kung CLA scales approximately as $\log_2(N)+3$ phases.}

\label{fig:phase_scaling_adders}
\end{figure}

The phase-scaling result in Fig.~\ref{fig:phase_scaling_adders} further shows the architectural advantage of the CLA. The RCA requires approximately one adiabatic phase per bit because the carry must propagate sequentially through the adder. In contrast, the Brent--Kung CLA computes the carries through a prefix structure, so the number of required phases grows much more slowly with bit width. For this PFAL implementation, the CLA requires 5 phases for 4 bits, 6 phases for 8 bits, and 7 phases for 16 bits, following the approximate scaling $N_{\mathrm{phases}}=\log_2(N)+3$. Therefore, the CLA is more suitable for scaling PFAL arithmetic.

The results show that, for this PFAL implementation, the number of gates alone is not the dominant factor. The complexity and capacitance of each individual PFAL gate are also critical, because they determine how effectively the power-clock can charge, hold, and recover the latch nodes. The CLA performs better than the RCA because it reduces the long carry-propagation limitation while using smaller PFAL cells that remain easier to drive.

\section{Conclusion}\label{sec:conclusion}

This work investigated how PFAL can be designed, optimized, and scaled for energy-efficient operation in TSMC 16\,nm FinFET technology. The results support the hypothesis that waveform-optimized PFAL can achieve lower energy than equivalent static CMOS when the power-clock waveform and circuit architecture are matched to the gate complexity. A functionally complete PFAL gate library was designed and verified using Buffer/NOT, AND/NAND, OR/NOR, and XOR/XNOR gates, providing the basis for larger combinational circuits.

At the single-gate level, PFAL showed clear optimum operating regions over voltage and frequency. The optimum EDP regions occurred in the GHz range, showing that PFAL operation in 16\,nm FinFET is not limited to low-frequency adiabatic switching. The Buffer/NOT and AND/NAND gates achieved better energy-delay behaviour than the XOR/XNOR gate, confirming that gate complexity, switched capacitance, and charging-path resistance strongly affect PFAL efficiency. The power-clock sweep showed that the minimum-energy waveform approaches a triangular shape rather than a conventional long-hold trapezoid. Sinusoidal excitation was then evaluated as a practical waveform and improved energy by up to $1.32\times$ compared with the trapezoidal reference, while also extending the valid voltage-frequency range.

At the combinational-circuit level, the 2:1 multiplexer, 4-bit RCA, and 4-bit Brent--Kung CLA showed that PFAL can scale from basic gates to arithmetic building blocks. The CLA was more suitable than the RCA, since it reduced carry-propagation depth and used smaller PFAL cells that were easier to drive with the power-clock. The 4-bit Brent--Kung CLA reached a gain of up to $5.3\times$ compared with the architecture-matched static CMOS energy estimate for the triangular power-clock.

Overall, the research question is answered. PFAL can be energy-competitive in 16\,nm FinFET technology when waveform selection, phase alignment, and circuit architecture are treated as coupled design choices. The strongest results are the single-gate energy gain of up to $3.83\times$ and the arithmetic-circuit gain of $5.3\times$ for the 4-bit Brent--Kung CLA, establishing a quantitative basis for waveform-optimized PFAL design.

\backmatter

\bibliography{sn-bibliography}

\end{document}